\documentclass[a4paper,english,aps,prb,twocolumn,amsfonts,amssymb,superscriptaddress]{revtex4} 
\usepackage[T1]{fontenc}
\usepackage[latin1]{inputenc}
\usepackage{graphics}
\usepackage{amssymb}
\usepackage{amsmath}
\usepackage{overpic}

\newcommand{\be}{\begin{equation}}
\newcommand{\ee}{\end{equation}}
\newcommand{\bea}{\begin{eqnarray}}
\newcommand{\eea}{\end{eqnarray}}

\def\a{\alpha}
\def\b{\beta}
\def\e{\varepsilon}
\def\d{\delta}
\def\g{\gamma}

\def\l{\lambda}

\def\n{\nu}
\def\o{\omega}

\def\s{\sigma}

\def\G{\Gamma}
\def\D{\Delta}

\def\ra{\rightarrow}
\def\up{\uparrow}
\def\pll{\parallel}
\def\down{\downarrow}

\def\pd{\partial}

\def\bk{{\bf k}}

\def\bq{{\bf q}}

\def\bv{{\bf v}}
\def\bQ{{\bf Q}}

\def\nn{\nonumber}
\def\lb{\label}
\def\pref#1{(\ref{#1})}

\newcount\bozza \bozza=0
\ifnum\bozza=1
\newdimen\shift \shift=-2truecm
\def\lb#1{%
{\label{#1}\rlap{\kern\shift{$\scriptstyle#1$}}}}
\else\def\lb#1{\label{#1}} \fi

\begin{document} 
\title{Anisotropy of the superconducting fluctuations in multiband
  superconductors:\\ the case of LiFeAs}

\author{L. Fanfarillo}
\affiliation{Instituto de Ciencia de Materiales de Madrid, ICMM-CSIC,
  \\
Cantoblanco, E-28049 Madrid, Spain} 
\author{L. Benfatto}

\affiliation{CNR-ISC and Dipartimento di Fisica, ``Sapienza''
  University of Rome, \\Piazzale A. Moro 2, 00185, Rome, Italy}

\begin{abstract}
  Between the different families of pnictide multiband
  superconductors, LiFeAs is probably one of the less
  understood. Indeed, despite the large amount of experiments
  performed in the last few years on this material, no consensus has
  been reached yet on the possible pairing mechanism at play in this
  system. Here we focus on the precursor effects of
  superconductivity visible in the transport experiments performed
  above $T_c$. By analyzing the superconducting fluctuations in a
  layered multiband model appropriate for this material, we argue that
  the strong two-dimensional character of the paraconductivity above
  $T_c$ points towards a significant modulation of the pairing
  interactions along the $z$ direction. We also discuss the peculiar
  differences between single-band and multi-band superconductors for
  what concerns the anisotropy of the superconducting-fluctuations
  effects above and below $T_c$.

\end{abstract}
%\pacs{74.20.Rp, 74.20.Fg, 74.25.Jb} 
 
\date{\today} 
\maketitle 

\section{Introduction}
After the original discovery of superconductivity in LaOFeAs\cite{kamihara} the
investigation of iron-based superconductors has lead to the discovery of several
classes of compounds that all display a high-temperature superconductivity,
despite the fact that the different lattice structures can lead to significant
differences in the electronic properties\cite{paglione, steward}. Such
differences justify the on-going debate on the existence of an universal pairing
mechanism in all iron-based superconductors. One of the crucial questions
concerns the role of antiferromagnetic spin fluctuations. Indeed, while they
represent a plausible candidate for the pairing glue in those systems, as 122
compounds, with good nesting conditions between the hole and electron Fermi
pockets\cite{steward,chubukov_review,reviewgap}, 
their relevance in other systems with poor nesting properties is often
questioned\cite{kontani,brydon_prb11,fete, kontani_cm14}. 
A typical example of such a system is LiFeAs. This compound is a
stochiometric superconductor with a $T_c\sim 17$
K\cite{pitcher_chem} and no magnetic ordering. Due to its stochiometric
nature and its clean, charge neutral cleaved surface LiFeAs is the best candidate
to perform both bulk- and surface-sensitive measurements. However, the
large amount of experimental findings accumulated so far did not
succeeded yet to clarify the nature of the pairing mechanism, but offered
instead a puzzling and somehow contradictory scenario. 
The Fermi surface (FS) of LiFeAs observed by angle-resolved
photoemission spectroscopy (ARPES) consists of two hole pockets at
$\G$ and two electron pockets at $M$\cite{borisenko_prl10, umezawa,kimura_prb12}, 
as in all pnictides families. However, there are some
remarkable quantitative differences, which are only partly accounted for
by LDA+DMFT (Dynamical Mean Field Theory) calculations\cite{valenti_prb12, arpes+dmft}.
In particular, the FS nesting is relatively poor and the hole pockets are
much shallower than in other families, with associated larger density
of states (DOS) which has been suggested to promote ferromagnetic
fluctuations instead of the antiferromagnetic
ones\cite{brydon_prb11}.
Nonetheless, as stressed by several authors, despite the poor nesting, a
spin-fluctuation mediated
pairing\cite{platt_prb11,arpes+dmft,hirschfeld_prb13,ummarino_13}, or
more generically a strong interband nature of the
interaction\cite{chubukov_prb14}, cannot be excluded. Indeed,
while perfect nesting is required to have a long-ranged SDW instability (absent
in LiFeAs), the presence of large spin fluctuations is sufficient to justify a
spin-mediated pairing mechanism, leading to a $s^\pm$ order parameter. Such
large magnetic fluctuations in LiFeAs have been observed by nuclear magnetic
resonance\cite{ma_prb10} and neutron scattering
experiments\cite{taylor_prb11,qureshi_prl12}, that also identified  a magnetic
vector slightly incommensurate\cite{qureshi_prl12} with respect to the
$\bQ=(\pi,\pi)$ wavevector, due probably to the bad nesting condition of the FS.
While these finding can support a $s^\pm$ scenario, its agreement with the ARPES
data on the gap modulation is still under
scrutiny\cite{umezawa,borisenko_prb11,borisenko_symm12}. Also the quasiparticle
interference patterns observed in scanning tunneling
microscopy\cite{allan_science12,hess_prl12,hess_cm12} are subject of an intense
debate. In this case different assumptions on the impurity scattering mechanism
can make the experimental results compatible either with a
$s^\pm$\cite{allan_science12} or with a (triplet)
$p$-wave\cite{hess_prl12,hess_cm12} symmetry of the order parameter.

The anomalies of LiFeAs are not restricted to the superconducting (SC) 
state but extend up to the so called normal state. 
The phe\-no\-me\-no\-lo\-gy and in particular the dimensionality of SC fluctuations 
above $T_c$ is an highly debated issue for all pnictides\cite{dim_SST14}.
However the situation in LiFeAs is particularly puzzling.
Indeed, while electronic properties of LiFeAs are believed to have an almost
three-dimensional (3D) character\cite{valenti_prb12}, as confirmed by de Haas
van Alphen experiments\cite{coldea_prl12,balicas_prb13}, the
superconducting fluctuations (SCF) exhibit a
marked two-dimensional (2D) fluctuation regime, which extends up to temperatures
very near to $T_c$\cite{song_epl12,rullier}. At the same time, the measurements 
of the upper critical field\cite{critical} slight below $T_c$ would point 
instead to a small anisotropy between in-plane and out-of-plane SCF, 
at odd with paraconductivity results. 
As we discuss in the present manuscript, the apparent contradictions between
these results can be reconciled by taking into account both the multiband
structure of pnictides and the peculiar interband character of the interactions.
In particular,  in multiband systems the link between the band structure and the
nature of SCF, both above and below $T_c$, can be more involved than what
expected in the single-band case, as pointed out in different contexts in the
recent literature.\cite{fanfarillo, gurevich_prb10,shanenko_prb13,marciani}
At the same time the interband character of the pairing can lead to remarkable
qualitative differences with respect to single-band systems, as it has been
emphasized in the context of transport\cite{fanfarillo_hall} and optical
properties above $T_c$\cite{benfatto_optical}.  In the present work we analyze
the SCF above $T_c$ in LiFeAs by taking the point of view of a spin-mediated
interband pairing mechanism, whose properties can be deduced by the analysis of
the SCF themselves. We show that the microscopic estimate of the crossover
temperature from 2D to 3D regime for the SCF is controlled by three cooperative
mechanisms: (i) the interband nature of the pairing, which leads to a weighted
contribution of the various bands in the single collective mode which controls
the critical SCF\cite{fanfarillo}; (ii) the low-energy renormalization effects
beyond Density Functional Theory (DFT), due the exchange of spin fluctuations\cite{benfatto_optical}; (iii) the
anisotropy of the pairing, that can make the SCF quasi-2D even for quasi-3D band
dispersions. On this respect our work supports recent theoretical
attempts \cite{hirschfeld_prb13,chubukov_prb14} to reproduce the measured gap hierarchy by taking into
account the possible variations of the pairing interaction along $z$,
due to the evolution of the FS. 
Our main finding is that the marked 2D character of SCF points
towards a prevalent 2D nature of the spin-fluctuation mediated pairing
interaction, that seems consistent with experimental observation of the magnetic
fluctuations above $T_c$\cite{ma_prb10,song_epl12}. Such a result is not
inconsistent with other estimates of the SC-properties anisotropy done below
$T_c$ with different probes, as the upper critical field\cite{critical}, the
thermal conductivity\cite{taillefer} and the critical current\cite{prozorov}.
Indeed, as we discuss below, in a multiband superconductor the weight of the
various bands to the SCF depends on the quantity under scrutiny, leading to
different results in the various experimental set-up. Finally, while our
findings cannot exclude an alternative pairing scenario based either on
ferromagnetic\cite{brydon_prb11} or orbital\cite{kontani_cm14} fluctuations, the
predominant {\em intraband} character of these mechanisms seems more difficult
to reconcile with an anisotropic pairing mechanism, crucial to interpret the SCF
above $T_c$.

\section{Collective critical mode}

Let us first of all summarize the expected result for the SCF anisotropy on
the basis of the derivation of Ref.\ [\onlinecite{fanfarillo}], that was done under
the following hypotheses: (i) the interaction has a predominant interband
character and (ii) the pairing is isotropic in momentum space (so it has
the same strength at all $k_z$ values). In this situation, it has been 
demonstrated that despite the presence of multiple FS 
the effective action of the SCF is still characterized by the emergence 
of a single critical collective mode that in the case of a layered 
superconductor is described by the propagator:
\be
\lb{defl}
L^{-1}(\bq,\o_m)=\nu \left[ \epsilon +\eta_\pll q_\pll^2+ r_z \sin(q_z d/2)+\gamma
|\o_m|\right]
\ee
where $\nu$ is the effective DOS of the collective mode at the Fermi level,
$\eta_{\parallel}$ and $r_z$ are the in-plane and the out-of-plane stiffness
respectively, $\e=\ln (T/T_c)$ and we used a periodic notation for the $q_z$
dispersion, with $d$ interlayer spacing. As a consequence the resulting
expression for the paraconductivity is the same obtained for a
single-band layered superconductor\cite{varlamov_book}, i.e.
\be
\lb{dslay}
\d \s=\frac{e^2}{16 \hbar d \sqrt{\e(\e+r_z)}}.
\ee
The crossover from 2D behavior $\d\sigma\sim 1/{\e}$ to 3D one
$\d\sigma\sim 1/\sqrt{\e}$ occurs at the temperature where $\e\lesssim
r_z$. In LiFeAs the 2D behavior is preserved until $\epsilon\sim
0.02$, so that one deduces that the out-of-plane stiffness $r_z$
is very small. In the single-band case the in-plane $\eta$
and out-of-plane $r_z$ stiffness can be estimated microscopically from
the values of the in-plane velocity and
out-of-plane hopping\cite{varlamov_book}:
\be
\lb{defeta}
\eta \sim \frac{v_F^2}{T^2}, \quad r_z\sim \frac{t_z^2}{T^2}.
\ee
In the multiband case the contribution of each band to the critical-mode values
for $\eta_\pll$ and $r_z$ depends in general on the relative strength of intra-
vs inter-band pairing\cite{varlamov_prb05,fanfarillo}. However, in the case of
pnictides the assumption of a predominant interband coupling simplifies
considerably the description of the critical collective mode. Following Ref.\
[\onlinecite{fanfarillo}] we shall consider a four-band model with only
interband pairing. By taking into account  DFT calculations\cite{valenti_prb12}
and ARPES evidences\cite{borisenko_prl10, umezawa} for LiFeAs we will
consider two electronic $\g_1,\g_2$ bands degenerate, and two hole bands $\a$ and $\b$,
corresponding to the inner and outer hole pockets, respectively. The larger
coupling $\l$ occurs between the quasi-nested $\a$ and $\g_i$ bands, while the
$\b-\g$ coupling $\l_{\b\g}=\kappa\l$, $\kappa<1$ is assumed to be smaller due
to the larger size of the $\b$ pocket. The BCS-like Hamiltonian of the model is
\be
\lb{ham}
H=\sum_i H_0^i+\lambda \sum_\bq \left[ \Phi^\dagger_{\g,\bq}
(\Phi^{\phantom{\dagger}}_{\a,\bq}+
\kappa \Phi^{\phantom{\dagger}}_{\b,\bq})+h.c.\right].
\ee
Here $H_0^i=\sum_\bk
\xi^i_{\bk}c^\dagger_{i,\bk\s}c^{\phantom{\dagger}}_{i,\bk\s}$,
$c_{i,\bk\s}^{(\dagger)}$ annihilates (creates) a fermion in the
$i=\a,\b,\g_1,\g_2$ band and $\xi^i_{\bk}$ is the layered 3D band dispersion
with respect to the chemical potential
\be
\lb{disp}
\xi^i_{\bk}=\frac{\bk_\pll^2}{2m_i}-t_{i,z}\cos (k_z d)-\mu.
\ee
In Eq.\ \pref{ham} $\Phi_{i,\bq}=\sum_\bk
c^{\phantom{\dagger}}_{i,\bk+\bq\up} c^{\phantom{\dagger}}_{i, \bk\down}$ is the
pairing operator in the $i$-th band, with
$\Phi_{\g,\bq}\equiv\Phi_{\g_1,\bq}+\Phi_{\g_2,\bq}$. It is possible to recast
the four-band model defined in Eq.\ \pref{ham} in an effective two-band model by
introducing the pairing operators  $\Phi_e\equiv\Phi_\g$ and
$\Phi_h\equiv\Phi_\a+\kappa\Phi_\b$ so that the pairing term reads:
\be
\lb{sa}
H_I=\lambda\sum_\bq(\Phi_1^\dagger\Phi_2^{\phantom{\dagger}}+h.c.).
\ee
Once established the pairing model according to Eq.\ \pref{sa}, we will use
band parameters consistent with LDA+DMFT and experimental measurements,\cite{valenti_prb12,
  arpes+dmft,umezawa,borisenko_prl10} and we will choose the interaction
strength in order to reproduce the experimental gap values. The estimate of
the fluctuation regime will then follow by the explicit calculation of the
critical multiband mode, done according to the analysis of
Ref. [\onlinecite{fanfarillo}]. 
Notice that despite the repulsive nature of the interaction \pref{sa}
a superconducting instability is still possible in the $s_\pm$ symmetry, where
the gap changes sign between hole and electron bands. However, in contrast to
the ordinary intraband-dominated pairing (as, e.g., in
MgB$_2$\cite{varlamov_prb05}) here a single pairing channel exists, with
important consequences on the implementation of the standard procedure to derive
the effective action for the SC fluctuations both above\cite{fanfarillo} and
below \cite{marciani} $T_c$. In particular, one can show that in the model
\pref{ham}-\pref{sa} the contribution of the various bands to the single
critical mode \pref{defl} is given by: 
\be
\lb{defc}
\eta_\pll=(w_h^2 \eta_{h}+w_e^2 \eta_{e}), \quad r_z=(w_h^2 r_{h}+w_e^2 r_{e})
\ee
where the coefficients $w_e$, $w_h$ are fixed by the two conditions:
\bea
w_ew_h&=&1,\\
\lb{ratio}
\frac{w_e^2}{w_h^2}&=&\frac{\D^2_e}{\D^2_h}=\frac{\Pi_h(q=0)}{\Pi_e(q=0)},
\eea
where $\Pi_h\equiv \Pi_\a+\kappa^2\Pi_\beta, \Pi_e=2\Pi_{\g_1}$ are the
Cooper particle-particle bubbles evaluated at zero frequency and momentum
($q=(i\o_n,\bq)$), $\D_h\equiv\D_\a=\D_\b/\kappa$ and $\D_e\equiv\D_{\g_i}$. The
last relation of Eq.\ \pref{ratio} has been derived from the usual
saddle-point equations $\D_e=-\lambda\Pi_h\D_h$ and
$\D_h=-\lambda\Pi_e\D_e$.  Analogously, the SCF parameters $\eta_{e(h)}$, $r_{e(h)}$, which are
obtained by the small $\bq$ expansion of the Cooper bubbles (see below), 
are given in term of the band stiffnesses as 
\bea
\lb{defetai}
\eta_e&=& 2 \eta_\g, \quad \quad \eta_h = \eta_\a + \kappa^2 \eta_\b,  \\
\lb{defri}
r_e &=& 2 r_\g, \quad \quad r_h= r_\a + \kappa^2 r_\b,
\eea
In first approximation the coefficients $\eta_i$ and $r_i$ can be extracted 
from the band parameters, listed in Table I, according to Eq.\
\pref{defeta}. For what concerns the relative weights $w_{e,h}$ in Eq.\
\pref{defc} one can extend the relation \pref{ratio} above $T_c$ where
$\Pi_i \simeq \nu_i ln(\o_0/T_c)$, i.e. each band is weighted inversely
proportional to its DOS $\nu_i$. By means of Eqs.\
\pref{defc}, \pref{ratio} and \pref{defri}, and by using the band
parameters listed in Table \ref{bp}, 
extracted from LDA+DMFT and experimental measurements,\cite{valenti_prb12,
  arpes+dmft,umezawa,borisenko_prl10} we can provide a preliminary
estimate of the anisotropy parameter $r_z$ for LiFeAs.  As one can see in
Table \ref{bp}, even though the outer hole band has $t_z\approx 0$, the
$t_z$ in the inner hole band and in the electron ones is quite larger, of
order of $15$ meV. As a consequence from Eq.\pref{defeta} one expects a
value $r_z\sim {\cal O} (10)$ even considering the weighting factors
$w_e,w_h$ defined in Eq.\ \pref{defc} to compute the average $r_z$ of the
critical collective mode.   Such an estimate can be hardly reconciled with
the experimental observation of a 2D regime for the SCF up to very small
$\e\sim0.02$, that would imply $r_z\sim 10^{-2}$. 

\begin{table}[h!]
\begin{center}
\begin{tabular}{cccccccccccccccc}
\hline \hline
 & & & && $\a$& & & &  &$\b$&  & & & &$\g_1,\g_2$\\
\hline
$m/m_e$  & & & & &4.51 & & & & & 5.86  & & & & &3.68 \\
\hline
$t$ (meV)  & & & & & 58 & & & & & 45& & & & & 72 \\
\hline
$\nu$ (eV)$^{-1}$ & & & & & 1.37 & & & & & 1.77 & & & & & 1.11\\
\hline
$\e^0$ (meV) & & & & &33 & & & & & 112 & & & & &-68 \\
\hline
$t_z$ (meV) & & & & & 15 & & & & & 0 & & & & & 17\\
\hline
$\Delta$ (meV) & & & & & 6.0 & & & & & 3.4 & & & & & 3.6\\
\hline
\hline
\end{tabular}
\caption{LiFeAs is a layered system (lattice parameters $a\sim3.9$
\, \AA, $d\sim6.5$ \, \AA). The relevant bands near the Fermi level $\a,\b,\g$
can all be approximated according to Eq.\pref{disp}. $m$ is
the in-plane mass, $t$ the in-plane hopping, $\nu$ the density of state, $t_z$ 
the out-of-plane hopping, $\epsilon^0= \e^e_{min}, \e^h_{max}$ the band edge and
$\D$ the gap. The band parameters and the gap values are extracted from
[\onlinecite{valenti_prb12, arpes+dmft, umezawa, borisenko_prl10}]. The
weighting factors $w_{e,h}$ in Eq.\ \pref{defc} are determined by the band
DOS, according to the relation \pref{ratio}. The gap values can be used
instead to tune the superconducting couplings $\l$ and $\l\kappa$, see Eq.\
\pref{sa}.} 
\label{bp}
\end{center}
\end{table} 

\section{Anisotropy of the pairing interaction}

\begin{figure*}[t]
\begin{center}
\includegraphics[clip=true,scale=.55,angle=0,clip=true]{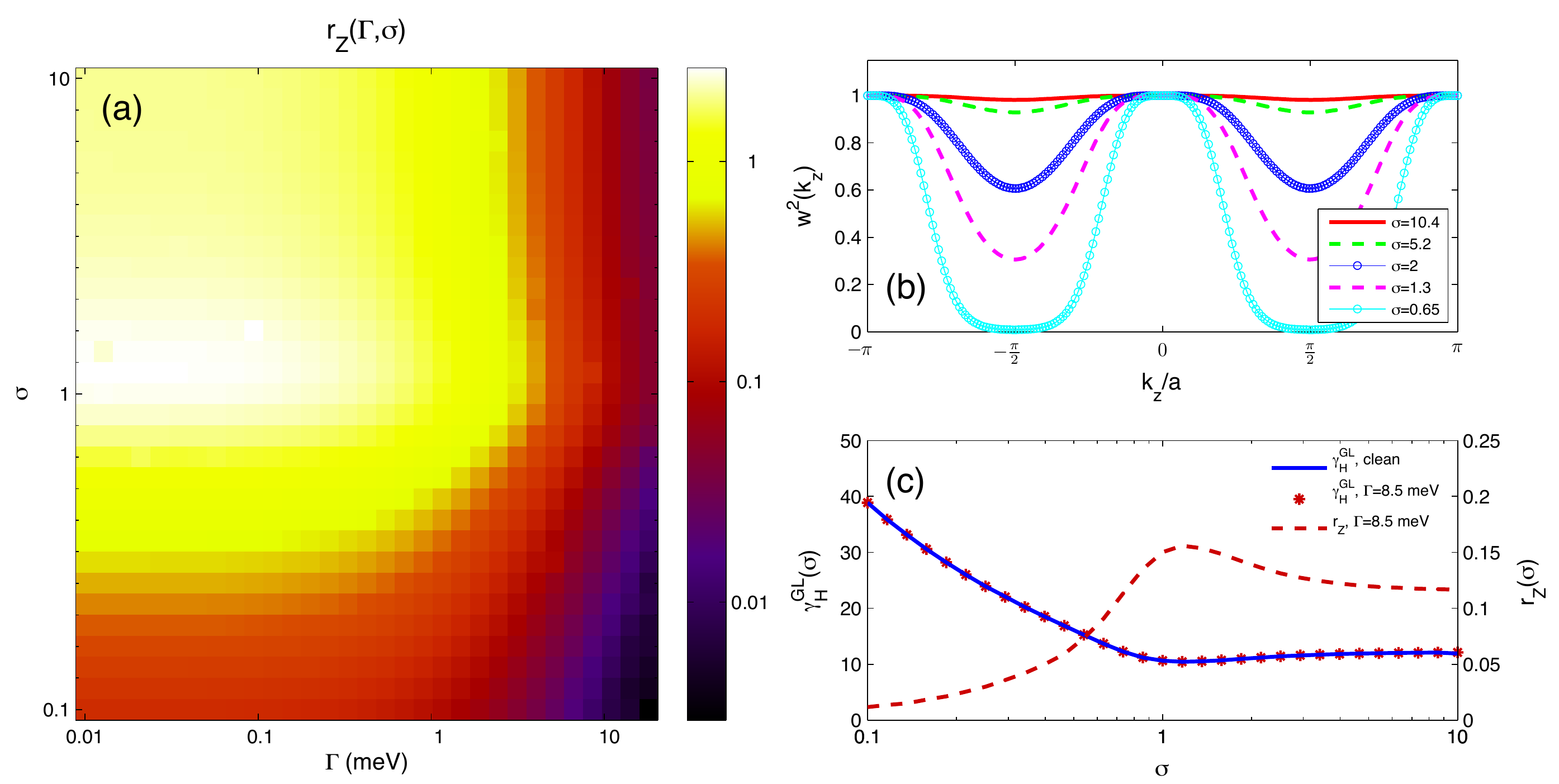}
\end{center}
\caption{(a) Dependence of the effective anisotropy parameters $r_z$ slightly
above $T_c$ ($T\sim 18$ K) on the scattering rate $\G$ and on the amplitude 
$\s$ of the $k_z$ weighting function $w(k_z)$ of Eq.\ \pref{wkz}. 
As $\G\ra 0$ meV and $\s \ra 10$ one recovers the result
of the clean, isotropic limit $r_z\sim 2$. 
By increasing the disorder, as well as squeezing the $w_\s(k_z)$
function, one finds a strong and sudden reduction of $r_z$.
(b) Parametric view of the $w^2_{\s}(k_z)$ function in the range of
integration in $k_z$. For $\s=10$ the weight $w_{\s=10}(k_z)\sim 1$ in the full
range of integration and one recovers the standard results. 
(c) Single-band estimate of the Ginzburg-Landau upper critical-field
anisotropy $\gamma_H^{GL}$, as  given by Eq.\ \pref{gamma_gl}, as a function of
$\s$.  The solid line and the
symbols correspond to the clean and dirty case ($\Gamma=8.5$ meV),
respectively. The value of $r_z$ for the same disorder level is also
reported (dashed line). } 
\label{plot}
\end{figure*}

All the above discussion has been based on the idea that the anisotropy of
SCF is simply determined by the anisotropy of the band structure, and band
parameters have been extracted from LDA+DMFT and ARPES measurements. In
this Section we discuss how this estimate can be modified by taking into
account several properties peculiar to pnictides. A first correction to be
considered is the low-energy band renormalization due to the same spin
fluctuations that mediate the pairing. Indeed, while LDA+DMFT correctly
accounts for the high-energy effects (like Hubbard-$U$ interactions) that
renormalize the overall bandwidth, spin fluctuations can give rise to an
additional band renormalization visible in a small energy range (of the
order of the spin-fluctuation scale $\o_0\sim 10-20$ meV) around the Fermi level. The dichotomy
between these two effects has been discussed for example in Ref.\
[\onlinecite{benfatto_optical}], where it has been show how these low-energy
renormalization effects are crucial to understand the discrepancy between
the effective masses probed by ARPES and the thermodynamical probes,
sensible to the carrier mass at the Fermi level. By using the results of an Eliashberg-like
approach to the spin-mediated interactions one can then introduce an
additional reduction of the hopping parameters listed in Table I as $t_i\ra
t_i/(1+\lambda)$, where $\l$ represents here an average dimensionless
coupling to spin fluctuations, estimated\cite{benfatto_optical} to be
in the intermediate-coupling regime $\l \sim 1-2$. By including this effect
we already reduce $r_z$ to $\simeq 2$, that is however still much larger
than the experimental value.
A second aspect to be considered is the modulation of the pairing interaction
along $k_z$. Indeed,  it is well known that in the case of
pnictides the structure of the pairing interaction along $k_z$ can be
definitively much more involved, as it has been recently pointed out in the 
case of P-doped BaFeAs \cite{nodez}.
By assuming that the pairing originates mainly from a spin-fluctuations
mediated interband mechanism, one must consider the evolution along the
$z$ direction both of the orbital character of the bands and of the
nesting properties between the (anisotropic) hole and electron pockets,
that contribute both to the effective $k_z$ dependence of the pairing
interaction. Both properties can vary between different materials and also
as a function of doping. For example, in Co-doped BaFeAs it has been
experimentally shown that spin fluctuations that are 3D anisotropic in the
undoped compound become much more 2D in the optimally-doped one, where a 2D
picture seems then more appropriate\cite{christianson}. 
Even though a detailed experimental investigation of this issue on LiFeAs is not yet
available, there have been already
suggestions\cite{hirschfeld_prb13,chubukov_prb14} for a possible $k_z$
dependence of the pairing interactions induced by the variations of 
the FS topology along $z$.
We analyze here the consequences on a anisotropic pairing interaction along $k_z$ 
for the properties of the SCF.  
In this case, while the structure \pref{defl}
of the critical collective mode does not change, we must reconsider the
estimate \pref{defeta} of the single-band parameters when the pairing has
an anisotropic structure along $k_z$. On very general
grounds\cite{varlamov_book,varlamov_prb05,fanfarillo} the SCF stiffnesses
$\eta_i, r_i$ are extracted from the small $\bq$ expansion of the Cooper
bubbles in each band. If one introduces explicitly a mo\-du\-la\-tion function
$w(k_z)$, that accounts for the variation of the pairing interaction along
the $z$-axis, the Cooper bubble is:
\bea
\Pi_i(\bq,0)&=&\frac{T}{N}\sum_{\bk,i\o_n}  w^2(k_z)G_i(\bk+\bq,\o_n)G_i(-\bk,-\o_n)=\nn\\
&=&\frac{1}{N}
\lb{defpi}
\sum_\bk
w^2(k_z)\frac{f(-\xi^{i}_\bk)-f(\xi^{i}_{\bk+\bq})}{\xi^{i}_\bk+\xi^{i}_{\bk+\bq}},
\eea
where $G_i$ is the Green's function of the $i$-th band above $T_c$.
By retaining leading terms in the $\bq^2$ expansion on Eq.\ \pref{defpi} one obtains:
\bea
\lb{def_eta}
\nu_i \eta_i&=& \nn \frac{1}{8N} \sum_{\bk} w^2(k_z) v^2_{i,\pll}(\bk)
\Bigg[\frac{f^\prime(\xi^i_\bk)}{(\xi^i_\bk)^2}+
\frac{\tanh{(\b\xi^i_\bk/2)}}{2(\xi^i_\bk)^3}\Bigg],\nn\\
\nu_i r_i&=& \frac{1}{4N} \sum_{\bk}w^2(k_z) v^2_{i,z}(\bk)
\Bigg[\frac{f^\prime(\xi^i_\bk)}{(\xi^i_\bk)^2}+
\frac{\tanh{(\b\xi^i_\bk/2)}}{2(\xi_\bk^i)^3}\Bigg],\nn\\
\lb{def_rz}
\eea
where $\bv_\bk=\pd \xi_\bk/\pd \bk$, so that for a band dispersion 
as in Eq.\ \pref{disp} $v_\pll(\bk)=\pd_{\bk_\pll}\xi_{\bk}$ and 
$v_z\sim \sin k_z$. 
The $\bk$-integrals in Eqs.\ \pref{def_eta}-\pref{def_rz} are dominated by
$\bk=(\bk_{\pll}, k_z)$ values at the FS. In 
particular for $w^2(k_z)=1$ one recovers the usual estimates \pref{defeta},
so that $r_z$ scales as $t_z$. However, when $w^2(k_z)$ is peaked at small
$k_z$ values, where $v_z\sim 0$, and it is reduced at intermediate $k_z d\simeq
\pi/2$, where $v_z$ is maximum, the effective out-of-plane parameter $r_z$ will be
strongly suppressed with respect to $t_z$. This effect is in part
compensated by an analogous reduction of the effective DOS $\nu_i$ that appears as a
prefactor in the expansion \pref{def_rz}, and that is now defined as:
\be
\lb{dos_weight}
\nu_i = \int_{d\xi_i}\d(\xi^i_F-\xi^i_\bk) w^2(k_z),
\ee
while the usual band DOS would be computed with $w^2(k_z)=1$. Since the anisotropy parameter scales as 
$r_z \sim \int dk_z v_z w^2(k_z)/ \int dk_z w^2(k_z)$ its overall reduction
is smaller than the one of the pairing-averaged out-of-plane velocity. 
In the following we will consider as a 
paradigmatic example a  modulation $w_\s(k_z)$ function defined as (Fig. \ref{plot}.b)
\be
\lb{wkz}
w^2_{\s}(k_z)= \exp\bigg[{-\frac{(1-\cos(2k_z))^2}{2\s^2}}\bigg],
\ee
and we will study the evolution of the effective anisotropy parameter $r_z$ as $\sigma$ changes.

Finally, to make a closer connection to experimental data we will also
account for disorder effects, that can be relevant in the regime of
temperature we are considering. Indeed while weak disorder
does not affect the $T_c$, defined by the $q=0$ limit of the Cooper
bubble, it modifies the stiffness\cite{varlamov_book}.  While for the
$q=0$ limit of the Cooper bubble the inclusion of vertex
corrections due to disorder is crucial, for an estimate of the
stiffness we can in first approximation use the bare-bubble scheme,
corresponding to replacing in Eq.\ \pref{defpi} the bare Green's function
with the one having a finite quasiparticle scattering rate $\G$, and
integrating over the frequency the corresponding broadened spectral
functions. As a consequence, the anisotropy coefficients of Eq.\
\pref{def_rz} are replaced by:
\bea
\n_i\eta_i&=&\frac{1}{4N}\sum_{\bk,k_z} w^2(k_z)v_{i,\pll}^2(\bk) \int dz\, dz'
A(\xi_i,z)A(\xi_i,z')R(z,z'),\nn\\
\n_i r_i&=&\frac{1}{2N}\sum_{\bk,k_z} w^2(k_z)v_{i,z}^2(\bk) \int dz\, dz'
A(\xi_i,z)A(\xi_i,z')R(z,z'),\nn
\eea
where $\xi_i$ is given by Eq.\ \pref{disp} and 
\bea
A(\xi,z)&=&\frac{1}{\pi}\frac{\G}{(z-\xi)^2+\G^2},\nn\\
R(z,z')&=&\frac{f'(z)+f'(-z')}{(z+z')^2}+2\frac{f(-z')-f(z)}{(z+z')^3}\nn.
\eea

The results for the  effective anisotropy
coefficient $r_z$ computed for LiFeAs are
shown in Fig.\ \ref{plot}.a. Here, to better clarify the interplay between the effects of disorder and of the 
anisotropy of the pairing interaction, we present a map of $r_z$ in terms of the scattering rate $\G$ and of the standard 
deviation $\s$ of the interaction's weight (see Fig. \ref{plot}.a)
The effective anisotropy parameter $r_z$ is maximum at $\G\ra 0$ meV and 
$\s \ra 10$, which represents the clean case and isotropic 
pairing interaction. 
Increasing the amount of disorder (i.e. increasing $\G$), as well as squeezing
the $w_\s(k_z)$ function by reducing its standard deviation $\s$, 
one observes a strong reduction of $r_z$. While a significant reduction of
$r_z$ can be obtained with these two cooperative mechanisms, the
experimental estimate of a $r_z$ as low as 0.02 would require a marked 2D
character of the pairing mechanism, along with a non-negligible residual
scattering rate $\G\simeq 10$ meV. While these estimates are not
inconsistent with the measured resistivity\cite{rullier} and with the 2D
character of the spin fluctuations above $T_c$\cite{ma_prb10,song_epl12},
explaining then paraconductivity experiments\cite{rullier,song_epl12}, the
comparison with the experimental findings below $T_c$, where the SC
properties appear rather isotropic, requires a
detailed discussion.

\section{Comparison with other experiments below $T_c$}

In the previous section we showed that the quasi-2D character of the
SCF above $T_c$ in LiFeAs can be understood by taking into account the
anisotropy of the pairing mechanism along the interlayer direction. A
crucial issue is then to compare this result with other estimates of
the SC-properties anisotropy done in the literature below $T_c$. Here
we discuss in details three experiments measuring the upper critical
field\cite{critical}, the critical current\cite{prozorov} and the
thermal conductivity\cite{taillefer}. In general, while comparing
paraconductivity experiments above $T_c$ with other probes below $T_c$,
two main differences due to the multiband nature of the system 
must be taken into account. First of all, for what concerns the connection
between the pairing mechanism and the SC gap below $T_c$, one should
consider that when the pairing is mediated by spin
fluctuations the Fermi-surface reconstruction
due to superconductivity below $T_c$ can reduce, within a self-consistent
scheme, the anisotropy of the pairing interaction. This implies for example
that the gap function below $T_c$ can be less anisotropic than what probed
by the SCF above $T_c$. Second, for what concerns more specifically the
behavior of the SCF, for a multiband superconductor the weighted contribution of
the various bands to the SCF is {\em not} the same for the different
experimental probes, in contrast to what happens in a single-band system.

Let us start with the estimate of the anisotropy 
$\g_H=H_{c2}^{\perp c}/H_{c2}^{\pll c}$ between the critical fields
perpendicular  and parallel to the $c$ axis,
respectively\cite{critical}. By converting the critical field in the
correlation length by using the standard formulas
$H_{c2}^{\pll c}=\Phi_0^2/2\pi(\xi^H_\pll)^2$ and
$H_{c2}^{\perp c}=\Phi_0^2/2\pi\xi^H_z\xi_\pll^H$ one has:
\be
\lb{ggl}
\g^{GL}_H=\frac{H_{c2}^{\perp c}}{H_{c2}^{\pll
    c}}=
\left(\frac{\xi^H_\pll}{\xi^H_z}\right)
\ee
In a standard single-band superconductor the correlation lengths
$\xi_\pll^H,\xi_x^H$ which enter the above formula coincide with the
ones obtained by the hydrodynamic expansion of the fluctuation
propagator \pref{defl}. Thus, one would estimate
\be
\lb{gamma_gl}
\g^{GL}_H=\left (\frac{4\eta_\pll}{r_z d^2}\right)^{1/2}
\ee
Notice that since $\gamma_{H}^{GL}$ is given by the ratio between the
in-plane and out-of-plane stiffness it is rather insensitive to
disorder. This is shown in Fig.\
\ref{plot}c, where we report the expected {\em single-band}-like
estimate \pref{gamma_gl} of $\gamma_{H}^{GL}$  for LiFeAs both in the clean and
the dirty case as a function of the pairing anisotropy.  We also show
in the same plot the anisotropy parameter $r_z$ for the dirty case. 
As one can see, when $r_z\simeq 0.02$, that would be
consistent with paraconductivity experiments above $T_c$, 
$\g^{GL}_H$ would be around 20, i.e. much larger than the
value $\g_H\sim 2.5$ obtained experimentally near $T_c$. 
However, in a multiband superconductor one
cannot in general identify the $\xi_\pll^H,\xi_z^H$ entering Eq.\
\pref{ggl} with the ones entering the paraconductivity {\em at zero
  magnetic field}. The reason is the following: the paraconductivity
is determined by the hydrodynamic expansion of the SC collective mode
which becomes critical at $T_c$ at zero magnetic field. Thus, as showed in Ref.\
[\onlinecite{fanfarillo}], one first diagonalizes the multiband problem at
zero frequency and momentum to 
identify the contribution of the various
bands to the critical SC mode. Afterwards one expands it at
small momenta, to obtain the propagator \pref{defl} which enters the
leading Aslamazov-Larkin diagrams contributing to the
paraconductivity. To solve instead the problem at finite
magnetic field one must diagonalize the multiband problem by
retaining gradient terms (i.e. the finite-momentum expansion)  in
the GL propagator for each band. This leads in general to the
identification of new multiband effective correlation
lengths $\xi_\pll^H,\xi_z^H$, where the various bands can contribute 
with different weights with respect to the zero-field case.
A typical example is the two-band case with only interband pairing. 
The in-plane correlation length entering the paraconductivity
can be deduced from Eq.\ \pref{defc}, while the upper critical field
$H_{c,2}^{\pll c}$ has been computed in Ref.[\onlinecite{gurevich_prb10}], 
and reads
\be
\lb{hc2}
H_{c,2}^{\pll}(T)=\frac{24 \pi \Phi_0 T_c (T_c-T)}{7\zeta(3)\hbar^2
  (v_e^2+v_h^2)/2}=\frac{\Phi_0}{2\pi (\xi_\pll^H)^2}
\ee
where $v_{e,h}$ are the velocities of the electron/hole bands,
respectively. As a consequence, since $\xi^2_\pll\propto \eta_\pll$
and $\eta_{e,h}\propto v_{e,h}^2$ from Eq.\ \pref{defeta}, we obtain:
\be
\xi^2_\pll\sim \frac{1}{2} (w_h^2 v_h^2+w_e^2 v_e^2), \quad 
(\xi^H_\pll)^2\sim \frac{1}{2} ( v_h^2+v_e^2).
\ee
This means in particular that the two bands
contribute equally to $\xi^H_\pll$, while this is not the case for
$\xi_\pll$. Notice also that the above estimate \pref{hc2} has been
done for two bands with the same mass anisotropy, which is not the
case for LiFeAs. Thus, the result for a three-band model as the one
used in Sec. II is not known yet, and no conclusions can be reached on
the expected values of $\xi_\pll^H,\xi_z^H$ in our case. We also note
in passing that recently the anisotropy of the correlation
length below $T_c$ has been inferred also by measurements of the critical current
at different magnetic fields\cite{prozorov}. Interestingly, these measurements show an
{\em increase} of the $\gamma_H$ ratio as the magnetic field decreases, with
variations near $T_c$ by about one order of magnitude between $H=0.5$ T
and $H\simeq H_{c2}$. This result could then 
could reconcile the apparent discrepancy between  the paraconductivity,
measured at zero field, and the
upper-critical fields results.

A second interest comparison can be done with the measurements of the thermal
conductivity reported in Ref.\ \cite{taillefer} Here it has been
shown that the thermal transport is quite isotropic in LiFeAs, both
for in-plane and out-of-plane heat current. This would rule out any
possible gap node or minima for the gap both within the $k_z=0$ plane
and along the $k_z$ direction. However, once more much care should be
used to interpret data in a multiband superconductor in terms of a
single-band scheme. In particular heat transport in a multiband
superconductor is dominated by the band with the smallest
gap\cite{golubov_prb11}. Thus, in LiFeAs one would expect a dominant
contribution coming from the $\b$ hole pocket, which is the less
interacting one and then less affected by the modulation of the
pairing mechanism proposed above. 

\section{Conclusions}
In summary, we studied a microscopic layered three-band model for the SCF in
LiFeAs. By using realistic band parameters, as extracted from the
experiments, we showed that the strong 2D character of the SCF found
experimentally can be understood as a signature of a strong anisotropy
along the inter-plane direction of the pairing interaction, which
compensates the low anisotropy of the band dispersion. While
within a single-band scenario it would be difficult to reconcile this
result with other measurements on the SC anisotropy below $Tc$, the
multiband character of pnictides makes such a comparison not
straightforward, leaving several questions open for future
investigation. 

\section{Acknowledgements} 
This work has been supported by the Italian MIUR  under the project
FIRB-HybridNanoDev-RBFR1236VV and the project PRIN-RIDEIRON-2012X3YFZ2,  and by the
Spanish Ministerio de Econom\'ia y Competitividad (MINECO) under the project
FIS2011-29680.

\end{document}